\documentclass[referee]{raa}      
\usepackage{graphicx,times}
\usepackage{natbib}
\usepackage{amssymb,amsmath}
\usepackage{multirow}
\usepackage{ulem}
\usepackage{adjustbox}
\usepackage{lscape}
\usepackage{footnote}
\bibpunct{(}{)}{;}{a}{}{,}
\emergencystretch=\maxdimen
\begin{document}
\title{Investigation of the shortest period Am type eclipsing binary TYC 6408-989-1}

\volnopage{Vol.0 (200x) No.0, 000--000}      
 \setcounter{page}{1}          

\author{Xiao-man Tian\inst{1, 2}}

\institute{
School of Aeronautics, Shandong Jiaotong University, 5001 Haitang Road, Changqing District, 250000 Jinan, China; {\it txmjlx2018@163.com} \\
\and
Center for Astronomical Mega-Science, Chinese Academy of Sciences 20A Datun Road, Chaoyang District, Beijing, 100012, P. R. China \\
              }

\vs\no
   {\small Received~~20xx month day; accepted~~20xx~~month day}

\abstract{
The first BV bands photometric observations and the low-resolution spectrum of the shortest period Am type eclipsing binary TYC 6408-989-1 have been obtained. The stellar atmospheric parameters of the primary star were obtained through the spectral fitting as follows: $T_{eff}=6990\pm117 K$, $\log g=4.25\pm0.26 cm/s^2$, $[Fe/H]=-0.45\pm0.03 dex$. The original spectra obtained by European Southern Observatory (ESO) were processed with IRAF package by us. Based on the ESO blue-violet spectra, TYC 6408-989-1 was concluded as a marginal Am (Am:) star with a spectral type of kA3hF1mA5 IV-V identified through the MKCLASS program. The observed light curves were analyzed through the Wilson-Devinney code. The final photometric solutions show that TYC 6408-989-1 is a marginal contact binary with a low mass ratio (q=0.27). The temperature of the secondary component derived through the light curve analysis is significantly higher than main sequence stars. In addition, TYC 6408-989-1 is a poor thermal contact binary. The temperature differences between the two components is about 1800K. TYC 6408-989-1 should be located in the oscillation stage predicted by the thermal relaxation oscillations theory (TRO) and will evolve into the shallow contact stage eventually. The very short period (less than one day), marginal Am peculiarity and quit large rotational velocity ($v\sin i \simeq 160 km s^{-1}$) make TYC 6408-989-1 became a challenge to the cut-off of rotation velocities and periods of Am stars. We have collected the well known eclipsing Am binaries with absolute parameters from the literature.
}

 \authorrunning{Tian}            
   \titlerunning{Investigation of the Am type eclipsing binary TYC 6408-989-1}  
\maketitle

\section{Introduction}
Among the chemically peculiar (CP) stars, metallic-line A-type stars (Am stars for short) attract the attention of the astronomical researchers, because of their extremely strong or weak element lines, high binary frequency and other characteristics. Almost all Am stars are A or F type stars. Am stars show following common characteristic: slow rotation compared with the normal A or F type stars (less than 120 $km s^{-1}$)\citep{1973ApJ...182..809A}; the spectra will show some remarkably characteristics as follows: weaker Ca II, K line  \citep{1940ApJ....92..256T,1948ApJ...107..107R} but enhanced Sr II line; the scandium and calcium elements are under-abundant while the iron-group elements, Y, Ba, and the rare earth elements are over-abundant \citep{1970PASP...82..781C}. The researchers found that almost all Am stars (more than $90\%$) are the components of binaries \citep{1961ApJS....6...37A,1965ApJS...11..429A,2010ASPC..435..257H}. It is generally accepted that the largest rotational velocity of Am stars is about $120 km s^{-1}$ \citep{1973ApJ...182..809A,2000ApJ...544..933A}. There is a lack of very short periods ($P_{orb} < 1.2 days$) Am type binaries  \citep{1996A&A...313..523B}, and the general explanation is synchronism in such systems would force the primary to rotate faster than 120 $km s^{-1}$. Nevertheless, Am peculiarity may have no significant correlation with rotational velocity \citep{2004IAUS..224..209M,2005A&A...442..563M}. In addition, there are
some normal A0-A3 type slow rotators without reasonable explanation \citep{1973ApJ...182..809A,1978ApJS...37..371W,2007A&A...463..671R}. It is not clear yet whether slow rotation should be one necessary condition for the formation of an Am star, or slow rotation is an individual result of the Am phenomenon. It seems like the Am peculiarity can also depend on evolutionary status (or age)
\citep{2000A&A...354..216B,2004IAUS..224..209M,2005A&A...442..563M}, atmospheric parameters \citep{1998A&A...330..651K,2000A&AS..144..203H}or orbital elements in a binary system \citep{1996A&A...313..523B,1997A&A...326..655B,1998A&AS..128..497I,2004IAUS..224..749F} as well. In classical Am stars, the spectral types inferred from the Ca II K line (Sp(K)) are early than that inferred from the metal lines (Sp(m)) with about five or more spectral subclasses. While stars in which the difference of spectral
type ($\Delta=Sp(K)-Sp(m)$) is less ($1 \leq \Delta < 5$) are often referred to as ‘marginal Am’(or Am:) stars.  Many Am stars show anomalous luminosity effect, i.e. the luminosity criteria in certain spectral regions will indicate that the target should be a giant or even supergiant star, whereas in other regions the luminosity will indicate it as a dwarf or even lower luminosity star.

Some theories were presented to expound the origin of Am phenomenon. The radiative diffusion occurred in a strong magnetic field will likely lead to chemical peculiarities of Ap stars \citep{1970ApJ...160..641M}, while when the magnetic field is absent, the diffusion will caused Am phenomenon \citep{1971A&A....13..263W}. The slow rotation will further assists the diffusion to segregate the elements \citep{1983ApJ...269..239M}. The spin braking may be a efficient process for the Am phenomenon in close binaries, but it
is difficult to explain that phenomenon occurred in single stars. Furthermore, the significantly weak but observable magnetic fields were found rarely in Am stars (e.g. Sirius A \citep{2011A&A...532L..13P}, $\beta$ UMa and $\theta$ Leo \citep{2016A&A...586A..97B}). Taking the radiative acceleration and atomic diffusion into consideration, \citet{2000ApJ...529..338R} developed the stellar evolutionary models. These models can produced resemble abundance anomalies resemble to that of Am stars with much larger value. We should note that the standard stellar evolution theory, the accretion processes effects of interstellar and the circum stellar gases should all be taken into account for the investigate of the atmospheric chemical abundances of Am stars in the binary systems. As proposed, the accretion processes may strongly influence the surface abundances, such as the mass transfer from the evolved companion \citep{1965ApJ...142..423F,1989ApJ...345..998P}. \citet{1965ApJ...142..423F} proposed that the peculiar A and B type stars have evolved into an advanced phase, Such stars have evolved through the giant phase and returned to the vicinity of the main sequence. In other words, such a process has occurred in one component of the close binary, mass transfer has happened from the evolved component to the peculiar star.

Near contact binaries (NCBs) were defined as eclipsing binary with following common characteristics: continuous EB-type light variations, facing surface less than 0.1 orbital radius apart, short periods (less than one day) and one or two component at or near their Roche lobes (RLs) \citep{1994MmSAI..65...95S}. NCBs actually were classified into the following subclasses: semi-detached, marginal contact and marginal detached systems \citep{2006MNRAS.367..423Z}. The mass transfer is frequently happened in NCBs. NCBs have been significant targets to study the transition  between the tidal-locked detached stage and W UMa type overcontact stage, occurring in binary systems.

TYC 6408-989-1 (ASAS J235103-1904.5, NSVS 14636842, GSC 06408-00989) is an EB type Am type eclipsing binary with a period of 0.470796 day \citep{1991A&AS...89..429R,1971AJ.....76..338S,2009A&A...498..961R,2014A&A...564A..69S}.
The spectral type of the primary component is A4m \citep{2014A&A...564A..69S}. \citet{1971AJ.....76..338S} found some early type stars including this target near the south galactic pole. The period of TYC 6408-989-1 is shortest among the well known Am type eclipsing binaries \citep{2009A&A...498..961R,2012A&A...537A.117T,2015AJ....150..154T}.
\citet{2014A&A...564A..69S} presented the light curve of this target obtained by the Super WASP(SWASP) survey without photometric solutions. No detailed investigation on this target has been carry out.

The first $BV$ bands light curves and low resolution spectra of TYC 6408-989-1 were obtained in this study. The detailed light curves analysis and spectral fitting were presented. We collected all well known Am type eclipsing binary with absolute parameters from the literature. The relationships between these parameters are summarized. Furthermore, the evolutionary stage of these systems and the reason of the chemical peculiarities are discussed.

\section{OBSERVATION AND DATA REDUCTION}

\subsection{PHOTOMETRIC OBSERVATIONS}
\begin{figure}
  \begin{center}
   \includegraphics[width=12cm]{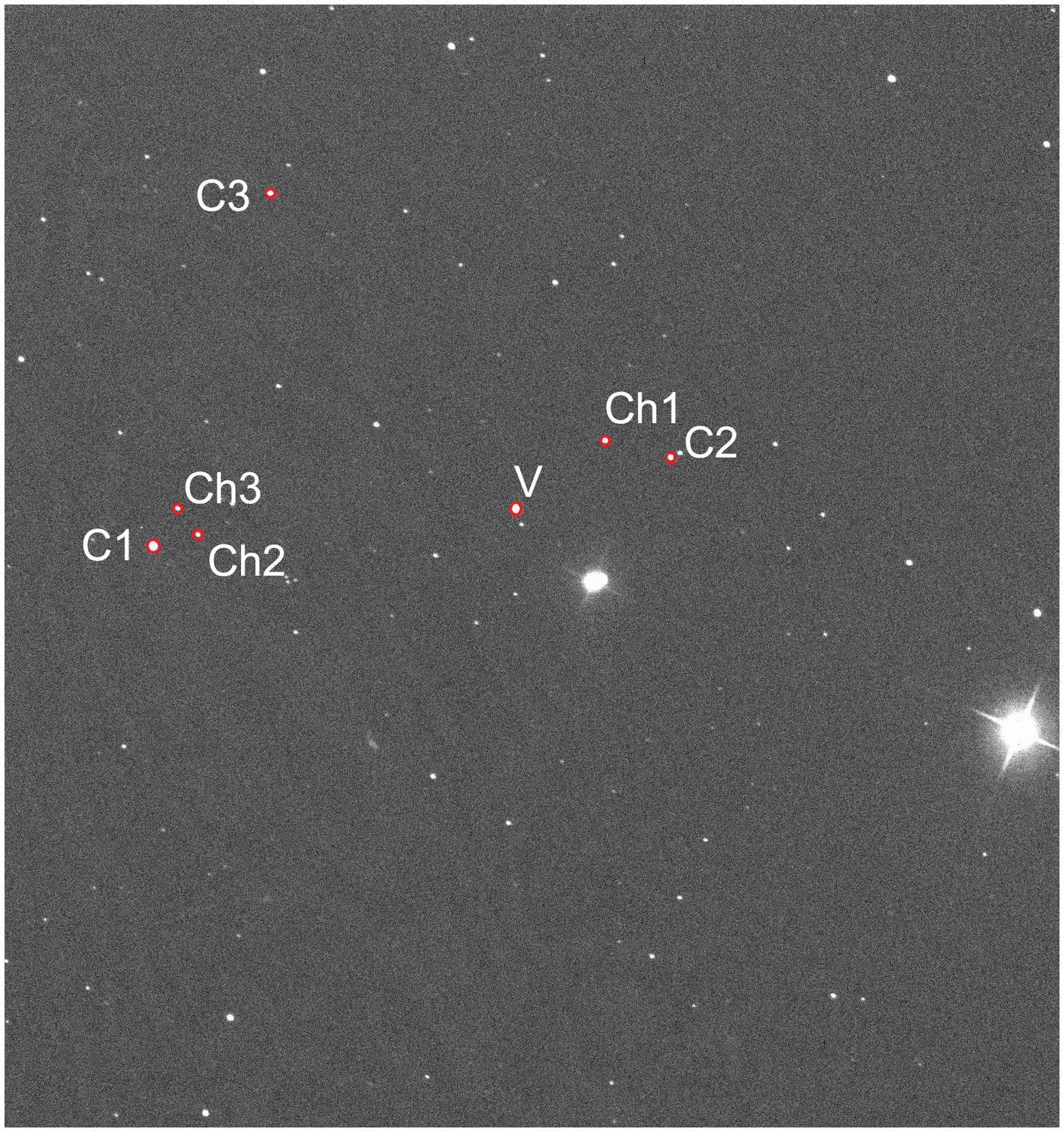}
  \end{center}
  \caption{Observed CCD image of TYC 6408-989-1. 'Variable star' was marked with 'V'. C1, C2, C3 marked the comparison stars and  Ch1, Ch2, Ch3 represented the check stars. }
\label{fig1}
\end{figure}

The first $BV$ bands light curves of Am type eclipsing binary TYC 6408-989-1 were observed with the 70 \emph{cm} Sino-Thai telescope at Lijiang Observing Station of Yunnan Observatories (YNO), Chinese Academy of Sciences (CAS) on 2015 December 9, 11, 13, 24. The PHOT package of IRAF was used to process all the observations. The comparison star and the check star were chosen to determine the differential magnitudes. The information of the comparison and check stars along with the target TYC 6408-989-1 are listed in Table \ref{table1}. In the table, the V band magnitude of variable star and the comparison star Ch1 were taken from \citet{2008AJ....136.1067H} and \citet{2014AJ....148...81M}, respectively. The G band magnitudes of other comparison stars and check stars were acquired from the Gaia DR2 data\citep{2018A&A...616A...1G}.
The observed CCD image of TYC 6408-989-1 was shown in Figure \ref{fig1}, in which the variable star was marked with 'V', and the comparison, check stars were marked as C1, C2, C3 and Ch1, Ch2, Ch3, respectively. The mean flux of comparison stars C1,C2,C3 (regarded as the comparison flux) and the mean flux of the check stars Ch1, Ch2, Ch3 (regarded as the check flux) were used to obtain the final differential magnitudes of TYC 6408-989-1, which show the light curves in Figure 2. In the figure, different colors marked the data observed on different nights. The magnitude differences between the comparison stars and the check stars are show in the bottom of the figure.

\begin{figure}
  \begin{center}
   \includegraphics[width=12cm]{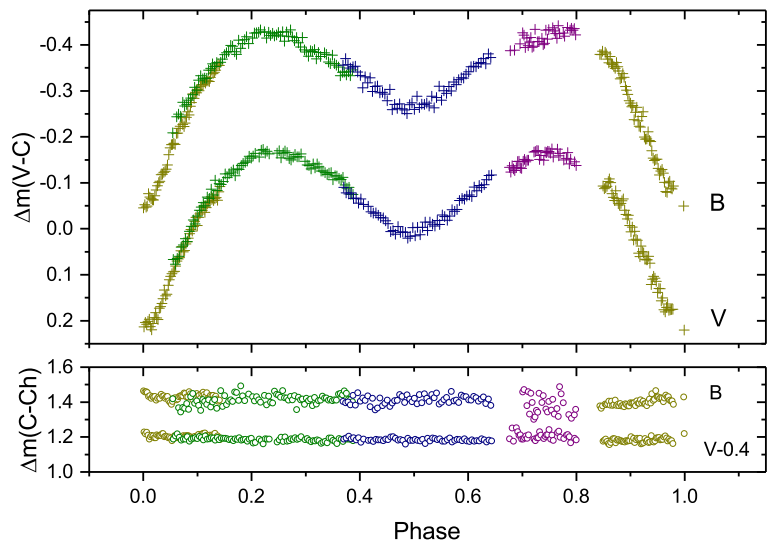}
  \end{center}
  \caption{Light curves of TYC 6408-989-1 on $BV$ bands observed with 70 \emph{cm} Sino-Thai telescope.}
\label{fig2}
\end{figure}

\begin{table}
  \caption{Information of TYC 6408-989-1, the comparison stars and the check stars.}
   \centering
\small
\label{table1}
\begin{tabular}{lclll}
\hline
 Targets & Name &${\alpha}_{2000}$  &  ${\delta}_{2000}$ & Mag    \\
\hline
Variable(V)  & TYC 6408-989-1& ${23}^{h}{51}^{m}{03}^{s}.5689$ & ${-19}^{\circ}{04}^{'}{29}^{''}.1076$ & V=12.064 \\
\hline
\multirow{3}{*}{Comparison(C)}
&C1(UCAC2 24500817) & ${23}^{h}{51}^{m}{06}^{s}.2789$   & ${-19}^{\circ}{11}^{'}{40}^{''}.3554$& V=12.390  \\
&C2(Gaia DR2 2390141539218676096)  & ${23}^{h}{50}^{m}{59}^{s}.6221$ & ${-19}^{\circ}{01}^{'}{24}^{''}.8279$ & G=13.8351 \\
&C3(Gaia DR2 2390132433888003200)&  ${23}^{h}{50}^{m}{38}^{s}.6209$ & ${-19}^{\circ}{09}^{'}{17}^{''}.6696$  & G=13.6923  \\
\hline
\multirow{3}{*}{Check(Ch)}
&Ch1(Gaia DR2 2390135560624200320) & ${23}^{h}{50}^{m}{58}^{s}.2628$ & $ {-19}^{\circ}{02}^{'}{42}^{''}.6827$ & G=14.0021  \\
&Ch2(Gaia DR2 2390130509742663808)& ${23}^{h}{51}^{m}{05}^{s}.3924$ & $ {-19}^{\circ}{10}^{'}{47}^{''}.3473$ & G=15.4127  \\
&Ch3(Gaia DR2 2390130475382925568)& ${23}^{h}{51}^{m}{03}^{s}.3342$ & $ {-19}^{\circ}{11}^{'}{11}^{''}.0596$ & G=14.966  \\
\hline
\end{tabular}
\end{table}

\subsection{SPECTRAL OBSERVATIONS}
\begin{figure}[htp!]
  \begin{center}
   \includegraphics[width=12cm]{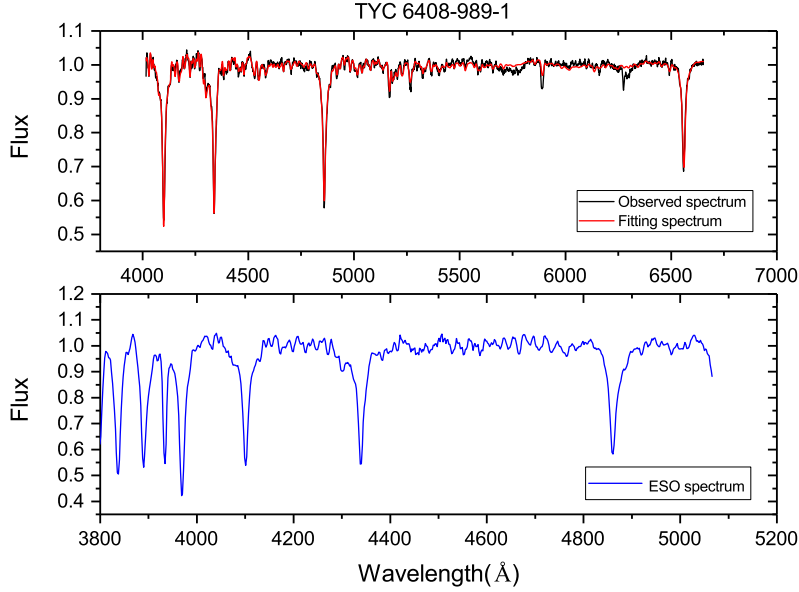}
  \end{center}
  \caption{Observed spectrum of TYC 6408-989-1. The black and red line in the upper panel represent the observed and fitted spectra, respectively. The blue line in the bottom panel shows an overview of the ESO spectrum.}
\label{fig3}
\end{figure}

The spectrum of TYC 6408-989-1 was observed on 2016 December 23 with the Beijing Faint Object Spectrograph and Camera (BFOSC) mounted to the 2.16 \emph{m} telescope of Xinglong station of National Astronomical Observatories of China (NAOC), Chinese Academy of Sciences (CAS). The low-dispersion spectrometer BFOSC and grating G7 were used during the observations. The slit width and line dispersion of grating G7 are 1.8 arcsec and 95 ${\AA} mm^{-1}$, respectively. The observable wavelength range is $4000-6800 {\AA}$. The observations process and spectra extraction were done using IRAF. The fluxes were normalized and the atmospheric absorption lines were corrected. In such a low resolution, the observed spectra only show the spectral lines of the primary component. The observed spectrum was shown in the upper of Figure\ref{fig3} with a black line. The University of Lyon Spectroscopic analysis Software (ULySS) \citep{2009A&A...501.1269K} was employed to acquire the atmospheric parameters through the full spectra fitting with the model spectra generated by an interpolator with the ELODIE library \citep{2001A&A...369.1048P}.
The fit spectrum was shown in the upper of Figure \ref{fig3} with a red line. The obtained atmospheric parameters are as follows: $T_{eff}\!=\!6990\!\pm\!117 K$, $\log g\!=\!4.25\!\pm\!0.26 cm/s^2$, $[Fe/H]\!=\!-0.55\!\pm\!0.03 dex$.
The primary star contributes the most light to the total system, the above atmospheric parameters were commonly applied to show the atmospheric characteristics of the primary star.

The spectroscopic data of TYC 6408-989-1 are available in European Southern Observatory (ESO) archives. The data were observed with the spectrometer Faint Object Spectrograph and Camera (EFOSC2) \citep{1984Msngr..38....9B} mounted at the New Technology Telescope (NTT) at La Silla Paranal observatory. During the observation, a 600 $g mm^{−1}$ grating (i.e. grism 14) was used. Grism 14 has a resolving power of 7.54 ${\AA}$ and a spectral range of $3095-5058$ ${\AA}$. The slit width and the dispersion of grism 14 are 1.0 $arcsec$ and 0.93 ${\AA} pixel^{-1}$. The observations process and spectra extraction were done by us using IRAF. An overview of the ESO spectrum was shown in the bottom of Figure \ref{fig3}. The ESO spectra were used to identify the spectra type through the MKCLASS program \citep{2014AJ....147...80G}, which was designed to classify stellar spectra on the MK Spectral Classification system in a way similar to humans—by direct comparison with the MK classification standards. The standard library $libr18$ was used in our classification progress, which consists of 1.8 ${\AA}$ resolution rectified spectra with a spectral range from $3800-4600 {\AA}$. The spectra in standard library $libr18$ were obtained with a 1200 $g mm^{−1}$ grating on the GM spectrograph of the Dark Sky Observatory (Appalachian State University).

\begin{figure}[htp!]
  \begin{center}
   \includegraphics[width=12cm]{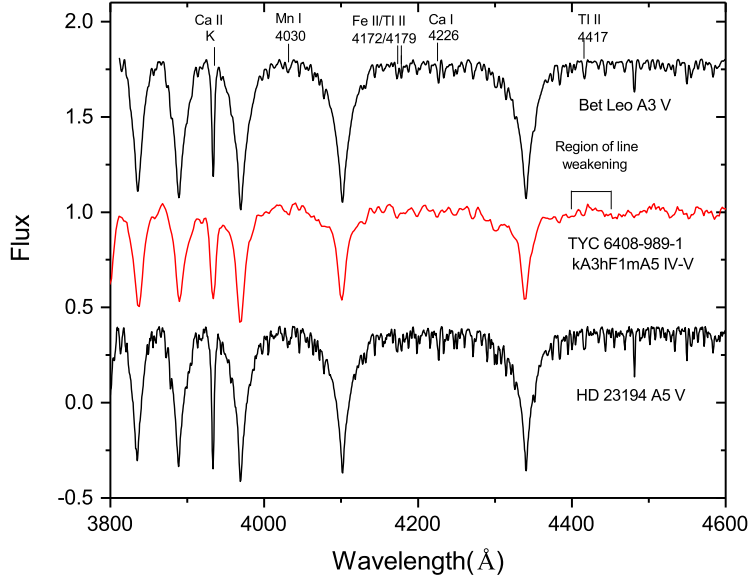}
  \end{center}
  \caption{A comparison of the blue-violet spectrum ($3800-4600$ ${\AA}$) of TYC 6408-989-1 (kA3hF0mA5 IV-V) and part of the spectrum of two MK standards Bet Leo (A3 V) and HD 23194 (A5 V). Note the difference in Ca II K line strength between TYC 6408-989-1 and the two MK standards.}
\label{fig4}
\end{figure}

A comparison of the blue-violet spectrum ($3800-4600$ ${\AA}$) of TYC 6408-989-1 (kA3hF1mA5 IV-V) and part of the spectrum of two MK standards Bet Leo (A3 V) and HD 23194 (A5 V) are shown in Figure \ref{fig4}. It is clearly that the Ca II K line of TYC 6408-989-1 is slightly weaker in strength than that of the A3V and A5V MK standards. The spectral type derived from the MKCLASS is kA3hF1mA5, and the index of spectral type difference $\Delta$ is 2. Therefore, we conclude that TYC 6408-989-1 is a marginal Am (Am:) star with a spectral type of kA3hF1mA5 IV-V.

\section{LIGHT CURVE ANALYSES}
The first $BV$ bands light curves of TYC 6408-989-1 were phased with following linear ephemeris:
\begin{eqnarray}
Min \emph{I}& = &HJD\,2457366.05772( .00037)+0.470796^{d}\times E
\end{eqnarray}
In the equation, the period was obtained from \citep{2014A&A...564A..69S}.
The Wilson-Devinney (W-D) program
\citep{1971ApJ...166..605W,1990ApJ...356..613W,2012AJ....144...73W} was used for the light curves analysis. The temperature of the primary component was estimated as $T_1=6990K$ through the spectral fit. The light curves with all modes of WD have been tried. The finally convergent solutions were achieved with mode 3 (over-contact mode). The gravity-darkening coefficients of both components are $g_1=g_2=0.32$ and bolometric albedos are $A_1=A_2=0.5$ \citep{1967ZA.....65...89L,1969AcA....19..245R}. The bandpass limb-darkening coefficients \citep{1993AJ....106.2096V} and the logarithmic bolometric coefficients were applied during the WD modelling.
There are some adjustable parameters in mode 3: the monochromatic luminosity of the primary star, $L_{1B}$, $L_{1V}$; the orbital inclination, i; the mean temperature of the secondary star, $T_2$; and the dimensionless potential of the primary star,$\Omega_1=\Omega_2$. The third light $L_3$ was an adjustable parameter, but no converged result was obtained. The value of the third light is negative and keeps decreasing, which means that there is no third light can be detected through the light curve analysis.

\begin{figure}[htp!]
 \begin{center}
  \includegraphics[width=12cm]{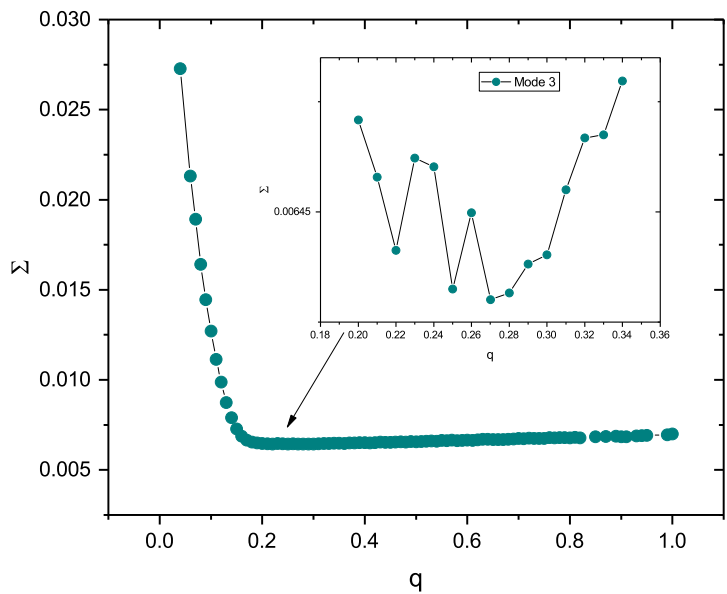}
  \end{center}
  \caption{$\sum - q$ curve of TYC 6408-989-1.}
  \label{fig5}
\end{figure}

\begin{figure}[htp!]
 \begin{center}
  \includegraphics[width=14cm]{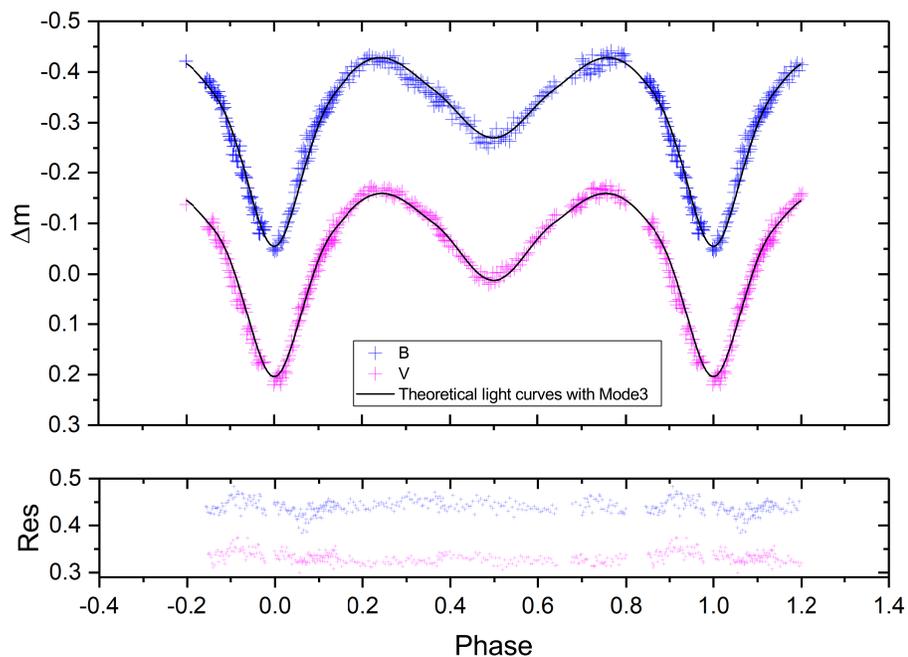}
  \end{center}
  \caption{The theoretical light curves of TYC 6408-989-1.}
  \label{fig6}
\end{figure}

\begin{figure}[htp!]
 \begin{center}
     \includegraphics[width=14cm]{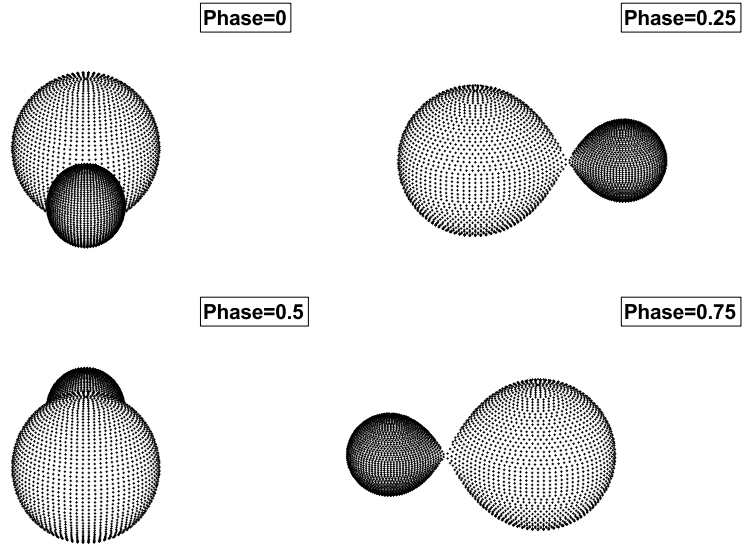}
  \end{center}
   \caption{Geometrical structure of TYC 6408-989-1 in 3D view.}
   \label{fig7}
\end{figure}

The mass ratio search (q-search) was a common technique to acquire the mass ratio.
The photometric solutions were obtained with some assumed mass ratio values from 0.01 to 1 with the differential correction program. The step of the search is 0.01. The relation between the sums of weighted square deviations ($\sum{(O-C)_i}^2$) and mass ratio ($q$) are shown in Figure \ref{fig5}. The minimal value achieved at $q=0.27$, which means that the solution at $q=0.27$ is the best fit. Then the mass ratio $q$ was adjustable. The final converged photometric solutions are listed in Table \ref{table2}. The uncertainty in the mass ratio in the table was given by the WD program with the standard method. In addition, the q-search curve bottom can be used to confirm the uncertainty of the mass ratio. It can be seen from Figure \ref{fig5}, the q-search bottom is clearly narrow around 0.270(3)(visually, from 0.25 to 0.28) and that the uncertainty is approximately 0.04. The theoretical light curves of TYC 6408-989-1 were shown in Figure \ref{fig6} with black lines. The standard deviations of the residuals on $BV$ bands both are about 0.012 magnitudes. The geometric structure in 3D view is plotted in Figure \ref{fig7}.

\begin{table}
\caption{Photometric solutions of TYC6408-989-1.}
\normalsize
\begin{center}
\label{table2}
\begin{tabular}{lc|lc}
\hline
Parameters       & Values& Parameters &Values  \\
\hline
Mode                             & Mode 3       &   ${L_{1}/L_{total}}_V$   & 0.93683(16)      \\
$g_{1}=g_{2}$                    & 0.32(fixed)  &   $f$                     & 0.003(29)  \\
$A_{1}=A_{2}$                    & 0.5(fixed)   &   $r_{1}(pole)$           & 0.46460(71)  \\
q ($M_2/M_1$ )                   & 0.270(3)   &   $r_{1}(side)$           & 0.50088(94)  \\
$i(^{\circ})$                    &  69.31(26)   &   $r_{1}(back)$           & 0.5255(11)  \\
$T_{1}(K)   $                    & 6990         &   $r_{2}(pole)$          & 0.2541(32)  \\
$T_{2}(K)$                       & 5089(50)     &   $r_{2}(side)$           &  0.2646(38)  \\
$\Omega_{1}$                     & 2.3985(49)   &   $r_{2}(back)$          &  0.2975(68)   \\
$\Omega_{2}$                     & 2.3985       &   $R_2/R_1$              & 0.5480(55)  \\
${L_{1}/L_{total}}_B$             & 0.959822(89)&   $\sum(O-C)^2$         & 0.0054          \\
\hline
\end{tabular}
\end{center}
\end{table}

\section{DISCUSSION AND CONCLUSION}
The $BV$ light curves of TYC 6408-989-1 were obtained firstly. The light curves show $\beta$ Lyrae characteristics. The atmospheric parameters of the primary star are estimated through the low resolution spectral fit as follows:  $T_{eff}\!=\!6990\!\pm\!117 K$, $\log g\!=\!4.25\!\pm\!0.26 cm/s^2$, $[Fe/H]\!=\!-0.55\!\pm\!0.03 dex$. The light curves analysis were done using WD program. The contact degree factor of TYC 6408-989-1 is 0.003(29), which is extremely low and the error is quite large. That means that TYC 6408-989-1 should be a marginal contact binary, in which both components has nearly filled their critical RLs, just like UX Eri \citep{2007AJ....134.1769Q}, AS Ser \citep{2008AJ....136..337Z}, DD Com \citep{2010AJ....140..215Z} and DI Hya \citep{2017PASP..129c4201L}.
TYC 6408-989-1 is one of the three well known Am type marginal contact binaries, the others are V1073 Cyg \citep{2018RAA....18...20T} and V2782 Ori \citep{2019PASP..131h4203T}. The orbital inclination of this target is $i=69.31(26)^\circ$. The mass ratio was estimated as about $q=0.270(3)$ through the mass ratio search progress. TYC 6408-989-1 was concluded as a marginal Am (Am:) star with a spectral type of kA3hF1mA5 IV-V. The difference between Sp(K) and Sp(M) is two subclasses ( i.e. $\Delta=2$). TYC 6408-989-1 consists of an F1 type primary star with $M_1=1.56(12)M_\odot$ \citep{2000PhT....53j..77C}. The spectra of the secondary star should be K0 type, which can be inferred from the temperature. Based on the photometric solutions, the absolute parameters of TYC 6408-989-1 were calculated as follows:
$M_2=0.42(10)M_\odot$, $R_1=1.581(3)R_\odot$, $R_2=0.874(2)R_\odot$, $L_1=5.360(378)L_\odot$, $L_2=0.459(3)L_\odot$. TYC 6408-989-1 should be a B sub-type contact binaries with temperature difference between the two stars (which is about 1860K) larger than 1000K \citep{1979ApJ...231..502L,2004A&A...426.1001C}, like V2787 Ori\citep{2019PASP..131h4203T}.

Using the Kepler's third law and the photometric solutions with mode 3, the mean densities of the primary and secondary star were calculated as $\rho_1=0.388\rho_\odot$ and $\rho_2=0.636\rho_\odot$, respectively. The corresponding logarithmic values are $\log \rho_1/\rho_\odot=-0.411$ and $\log \rho_2/\rho_\odot=-1.97$. The mean densities of both components (especially the secondary star) are significantly lower than main sequence stars with the same spectral type \citep{2000PhT....53j..77C}, which indicates that the components of TYC 6408-989-1 may have evolved away from the Zero Age Main Sequence (ZAMS) line and the secondary star may have a higher degree of evolution. The luminosity class of the target from IV to V indicates that the primary should be an evolved subgiant.

The components in very short period NCB systems like TYC 6408-989-1 are likely rotating synchronously. For a circular orbit, following equation can be used to obtain the synchronism rotational velocity:
$v = 50.6(R/R_\odot)/(P/d) km s^{-1}$\citep{2007MNRAS.380.1064C}, in which $R$ is the radius of the considered component, $P$ is the orbital period, $v$ is its equatorial rotational velocity. The synchronous rotation velocity $v \sin i$ of the primary  should be about 160 $km s^{-1}$. TYC 6408-989-1 is a very especial Am type eclipsing binary with very short period (less than one day), marginal Am peculiarity and quit large rotational velocity, which make it became a challenge to the cut-off of rotation velocities and periods of Am stars, like V1073 Cyg ($v \sin i = 150 km s^{-1}$)\citep{1996A&A...313..523B}.

According to the thermal relaxation oscillation theory \citep{1976ApJ...205..208L,1976ApJ...205..217F,1977MNRAS.179..359R,1979ApJ...231..502L}, W UMa type systems must undergo oscillations around the state of marginal contact. TYC 6408-989-1 should be a particular target like other marginal contact binaries lying in the TRO stage \citep{2002ApJ...568.1004Q,2009AJ....137.3574Z,2012AJ....144...37Z}
Such targets are very important for investigation of the formation between detached and contact phase of binaries, which are quite rarely in observation. The light curves of these systems show $\beta$ Lyrae-type light-variation characteristics. In the evolutionary process, the systems may be in a semidetached phase with the more massive star filling the RL, and then evolve into a marginal-contact phase with poor thermal contact. TYC 6408-989-1 should be lying on the poor thermal contact stage. It is another promising candidate locating in such rare evolutionary stage, it will evolve to become a shallow contact binary with true thermal contact(e.g.,\citet{2012AJ....144..178L,2013AJ....146...38Q, 2014ApJS..212....4Q} and \citet{2019PASP..131a4202L}).

\begin{figure}[htp!]
 \begin{center}
  \includegraphics[width=14cm]{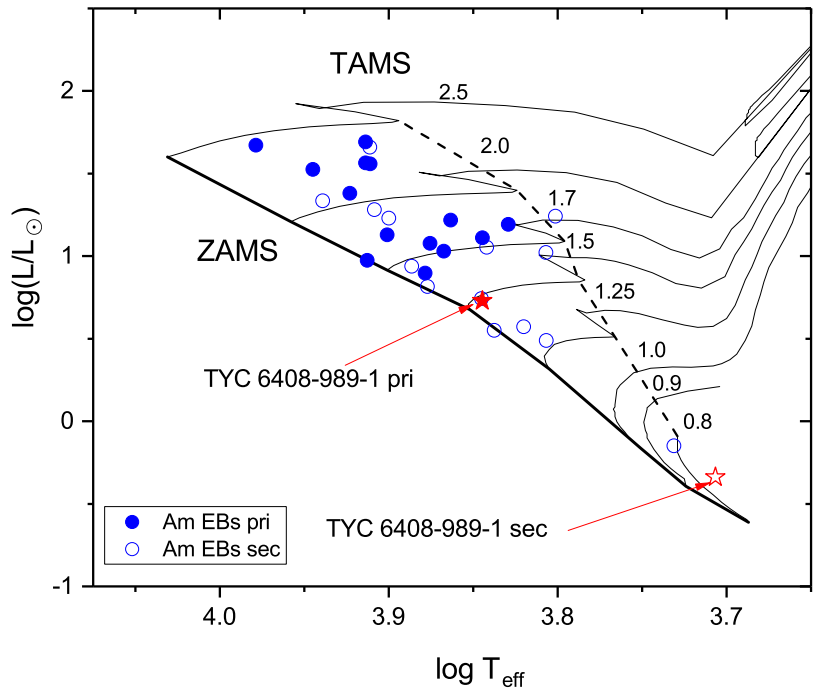}
  \end{center}
  \caption{The temperature-luminosity diagram of Am type EBs.}
  \label{fig8}
\end{figure}

Some eclipsing Am binaries with known parameters were collected and tabulated in Table \ref{table3}. The temperature-luminosity diagram of Am type EBs are plotted in Figure \ref{fig8}. In the figure, the ZAMS and terminal-age main sequence (TAMS) lines are obtained from \cite{1992A&AS...96..269S} at Z=0.020. From the figure, we can see that both components of Am type EBs are located in the main sequence. According to \citet{1965ApJ...142..423F}, one component in the Am type binary may has evolved through the giant phase, it may have striped most of the hydrogen shell and the stellar hell were exposed leading to higher temperature, then it will returned to main sequence. The statistical result support the above theory well. The accretion processes (such as mass transfer) may be very important for the formation of Am peculiarity in binaries. The mass transfer from the evolved component of Am type eclipsing binaries will recombine the chemical elements in the atmosphere of the companion stars and may result in chemical anomalies composition, such as the marginal contact binary V1073 Cyg \citep{2018RAA....18...20T}, V2787 Ori \citep{2019PASP..131h4203T} and TYC 6408-989-1, which are lying on the thermal relaxation oscillation state \citep{1976ApJ...205..208L,1977MNRAS.179..359R,1979ApJ...231..502L}. Mass transfer between the components may paly an important role in the formation and evolution of Am stars in binaries. More observation evidence are need to support this discussion. The researches on individual Am type binaries including the binaries with very short periods (less than one day) will be very important and indispensable for exploring the formation and evolution of Am star.

\begin{landscape}
\begin{table}[]
\centering
\caption{Parameter of some known Am type eclipsing binaries.}
\label{table3}
\begin{tabular}{llllllllllll}
\hline
Star           & Renson & Period     & q$(M_2/M_1)$ & $M_1$       & $M_2$       & $T_1$     & $T_2$     & $L_1$       & $L_2$       & Type        & Ref        \\
-              &        & (days)     &              & $(M_\odot)$ & $(M_\odot)$ & (K)       & (K)       & $(L_\odot)$ & $(L_\odot)$ &             &            \\
\hline
V1073 Cyg      & 56830  & 0.7858506  & 0.303        & 1.810(4)    & 0.549(1)    & 7300      & 6609(18)  & 16.51       & 3.74        & W UMa       & (1)        \\
V2787 Ori      & 11387  & 0.810979   & 0.120(2)     & 1.44        & 0.17(1)     & 6993(82)  & 5386(29)  & 12.93       & 0.71        & W UMa       & (2)        \\
TYC 6408-989-1 & 61280  & 0.470796   & 0.230(3)     & 1.56(12)      & 0.42(10)        & 6990      & 5089(50) )  & 5.36(38)        & 0.459(3)        & W UMa       & This study \\
TX Her         & 44140  & 2.05984    & 0.895        & 1.61        & 1.44        & 8180      & 7536(20   & 9.44        & 6.56        & Algol       & (3)        \\
YZ Cas         & 1140   & 4.4672235  & 0.585(2)     & 2.263(12)   & 1.325(7)    & 9520(120) & 6880(240) & 46.989      & 3.556       & Algol       & (4)(5)     \\
WW Aur         & 12320  & 2.52501941 & 0.9235(27)   & 1.964(7)    & 1.814(7)    & 7960(420) & 6411(410) & 13.458      & 10.544      & Algol       & (6)        \\
GZ CMa         & 15390  & 4.80085    & 0.909        & 2.200(25)   & 2.000(25)   & 8810      & 8694      & 33.496      & 21.627      & Algol       & (7)    \\
V624 Her       & 45460  & 3.894977   & 0.8326       & 2.270(14)   & 1.870(13)   & 8147      & 7943      & 36.308      & 16.982      & Algol       & (8)        \\
DV Boo         & 35946  & 3.782624   & 0.75         & 1.64(2)     & 1.23(2)     & 7370(80)  & 6410(74)  & 10.715      & 3.090       & Algol       & (9)        \\
V885 Cyg       & 50855  & 1.69478781 & 1.114        & 2.005(29)   & 2.234(26)   & 8375(150) & 8150(150) & 23.988      & 45.709      & Algol       & (10)       \\
EI Cep         & 57180  & 8.4393522  & 0.9483       & 1.7716(66)  & 1.6801(62)  & 6750(100) & 6950(100) & 15.596      & 11.350      & Algol       & (11)       \\
EE Peg         & 57410  & 2.628      & 0.6186       & 2.15(2)     & 1.33(1)     & 8709      & 6456      & --          & --          & Algol       & (12)       \\
SW CMa         & --     & 10.091988  & 0.9397       & 2.239(14)   & 2.104(18)   & 8200(150) & 8100(150) & 36.813      & 19.099      & Algol       & (13)       \\
HW CMa         & --     & 21.1178329 & 0.9665(43)   & 1.721(11)   & 1.781(12)   & 7560(150) & 7700(150) & 7.907       & 8.710       & Algol       & (13)       \\
AN And         & 60340  & 3.219566   & 0.53         & 2.48        & 1.32        & 8200      & 6330      & 49.2        & 17.5        & $\beta$ Lyr & (14)       \\
V501 Mon       & --     & 7.02       & 0.8865       & 1.6455(43)  & 1.4588(25)  & 7510(100) & 7000(90)  & 11.940      & 5.534       & Algol       & (15)       \\
\hline
\end{tabular}

\footnotesize{
\centering
(1)\citet{2018RAA....18...20T} (2)\citet{2019PASP..131h4203T}  (3)\citet{2019RAA....19...94Z}   (4)\citet{2014MNRAS.438..590P} (5)\citet{2000astro.ph.11388L}
(6)\citet{2005MNRAS.363..529S} (7)\citet{1985AJ.....90.1324P}  (8)\citet{1984AJ.....89.1057P}\\(9)\citet{2004MNRAS.352..708C} (10)\citet{2004AJ....128.1324L}
(11)\citet{2000AJ....119.1942T} (12)\citet{1984ApJ...281..268L}(13)\citet{2012A&A...537A.117T}
(14)\citet{1978JRASC..72..263T} (15)\citet{2015AJ....150..154T}
}
\end{table}
\end{landscape}

\begin{acknowledgements}
We greatly appreciate and feel great thankfulness for the hard work in the fight against COVID-19 to the Chinese people and all the people worldwide. We would like to thank the editor and the referee very much for the very valuable and useful comments that helped improve this paper. This work is partly supported by the National Natural Science Foundation of China (Nos. 11922306, U1831109), and the CAS Interdisciplinary Innovation Team. We acknowledge the support of the staff of the Xinglong 2.16 \emph{m} telescope. This work was partially supported by the Open Project Program of the Key Laboratory of Optical Astronomy, National Astronomical Observatories, Chinese Academy of Sciences. We acknowledge the kindly help of Pro. Gray Richard O.

CCD photometric observations of the system were obtained with the 70 \emph{cm} telescope at Lijiang station, Yunnan Observatories. The spectral observations were obtained with the 2.16 \emph{m} telescope at Xinglong station of the National Astronomical Observatory. The spectra classification were based on observations made with ESO Telescopes at the La Silla Paranal Observatory under programme 083.D-0472(A).
\end{acknowledgements}



\newpage
%

\end{document}